\documentclass[journal]{IEEEtran}
\usepackage{cite}
\usepackage{amsmath,amssymb,amsfonts}
\usepackage{algorithmic}
\usepackage{graphicx}
\usepackage{textcomp}
\usepackage{xcolor}
\usepackage[mathscr]{euscript}

 \usepackage{physics}

\usepackage[caption=false,font=footnotesize]{subfig}

\usepackage{cite}

\ifCLASSINFOpdf
\else
\fi

\begin{document}
%
\title{Joint Modelling of Quantum and Classical Noise over Unity Quantum Channel}
%
%
%


\author{Mouli~Chakraborty,~\IEEEmembership{Student~Member,~IEEE,}
       Harun~Siljak,~\IEEEmembership{Senior Member,~IEEE,}
      Indrakshi~Dey,~\IEEEmembership{Senior Member,~IEEE},
      and Nicola~Marchetti,~\IEEEmembership{Senior Member,~IEEE}
\thanks{M. Chakraborty is a PhD student at the Department of Electronic and Electrical Engineering,The University of Dublin,Trinity College, Dublin, Ireland; e-mail: chakrabm@tcd.ie.}
\thanks{ I. Dey is a professor at the Dept. of Electronic Engineering, Hamilton Institute Maynooth University, Ireland;  e-mail: indrakshi.dey@mu.ie}

\thanks{H. Siljak and N. Marchetti are professors at the Department of Electronic and Electrical Engineering,The University of Dublin,Trinity College, Dublin, Ireland; e-mail: harun.siljak@tcd.ie.; nicola.marchetti@tcd.ie }}
\maketitle

\begin{abstract}
For a continuous-input-continuous-output arbitrarily distributed quantum channel carrying classical information, the channel capacity can be computed in terms of the distribution of the channel envelope, received signal strength over a quantum propagation field and the noise spectral density. If the channel envelope is considered to be unity with unit received signal strength, the factor controlling the capacity is the \textit{noise}. Quantum channel carrying classical information will suffer from the combination of classical and quantum noise. Assuming additive Gaussian-distributed 
classical noise and Poisson-distributed quantum noise, we formulate a hybrid noise model by deriving a joint Gaussian-Poisson distribution in this letter. For the transmitted signal, we consider the mean of signal sample space instead of considering a particular distribution and study how the maximum mutual information varies over such mean value. Capacity is estimated by maximizing the mutual information over unity channel envelope.
\end{abstract}
\vspace{-3mm}   

%
\IEEEpeerreviewmaketitle

\section{Introduction}
%
%
%
%
\IEEEPARstart{H}{uman} beings can only generate and perceive classical information. In order to transmit classical information over a quantum communication channel, classical information is encoded into quantum states at the input of quantum channel. On the receiver side, the quantum states are decoded back to the classical information by measuring the output quantum states. Capacity of a classical information $-$ quantum channel $-$ classical information communication system can be calculated by maximizing the entropic measure of the transmitted classical information content per unit time. Understanding quantum communication from a classical transmission point of view is becoming even more important owing to the emerging promise of Quantum Internet and long-distance Quantum Satellite Networks \cite{1g} with near-zero latency and near-optimum security. This is because the control and management of a quantum network will rely heavily on classical networks, either bespoke or integrated as part of the existing network infrastructure. In future, we will design and deploy classical networks exchanging classical information between communicating nodes exploiting quantum technologies for realizing the communication links.

There is a rich body of literature on calculating quantum information capacities of quantum channel ranging from Shannon's quantum \cite{1a} to Holevo's capacity \cite{1b} and entanglement-assisted Holevo bound \cite{1c} for a Gaussian quantum channel carrying quantum information. However, these bounds consider the case where the channel is linear with Gaussian distributed input, output and noise. The quantum limit to classical communication considers only the classical environment
disregarding any quantum effects, while the quantum limit on quantum communication considers only the quantum environment disregarding classical channel effects.
However, in a realistic quantum communication network carrying classical information, one will encounter both classical and quantum channel effects and uncertainties.


Noise is one of the most challenging impairments affecting the processing of signals over which we have incomplete control. While it is possible to remove external noise, the internal noise caused by the spontaneous fluctuations of current or voltage within classical and quantum devices is very difficult  to remove, and therefore needs to be characterized in detail to gauge the scale of its impact. Internal noise can be broadly categorized as quantum shot noise (phononic noise) and classical thermal noise. Since the number of charged carriers (electrons) within a conductor is very large and their random motions are statistically independent of each other, the central limit theorem indicates that the thermal noise is Gaussian distributed with zero mean \cite{1d, 2d}. Quantum shot noise over the quantum channel can be visualized as as consequence of a large number of statistically independent interactions of the propagating signal with the quantum heat bath i.e. $Z (= Z_1 + Z_2 + \dotso + Z_N)$. The individual distribution of $Z_n$ can be one with zero mean and variance derived from the fluctuation-dissipation theorem, such that $Z$ can be approximated by the Poisson distribution \cite{1e, 2e}. 

In this letter, we model the hybrid quantum-classical noise by considering the summation of random variables and thereby, adding Poisson distributed quantum shot noise with Gaussian distributed white classical noise. Since we consider quantum communication from a classical transmission system perspective, we need to analyze the impact of a plethora of noise sources that arise for different entanglement and discord transmission scenarios (different memories and repeater technologies), while considering various classical signal coexistence scenarios. Though classical and quantum noises vary widely in temporal scales, it is possible to map the field and intensity distributions of the classical input signal to the amplitude and the probability distributions of quantum states and vice-versa, as in demonstrated in \cite{5a, 5b}. Stemming from these observations, we draw equivalence between the qualitative behaviors of quantum and classical noises (uncertainties) only in terms of their probability distributions; not in terms of evolution of the two noise processes in spatial and temporal scales.

The primary contribution of this paper is two-fold. Firstly, we derive a joint model for hybrid quantum-classical noise. Secondly, we calculate the mutual information between input and output by considering the mean input classical signal, and maximize the mutual information to calculate achievable capacity. 
Assuming unit channel envelope, numerical results reveal that classical-quantum-classical system exhibits increase in channel capacity with received SNR; a trend agreeing with classical communication theory, thereby establishing the validity of our mathematical treatise. In future, this joint noise model can be extended to calculate noise spectral density and, in turn, can be used to calculate channel capacity for arbitrarily distributed received signal envelope, as is done in \cite{1f}.

\section{System and Signal Model}
We considered a communication system where both source and destination are classical in nature. The source generates classical information which is encoded into quantum states. These quantum states are communicated over a quantum communication link (or a qubit channel). At the destination, classical information is retrieved back by measuring the quantum states obtained at the output of the link. In this case, the quantum channel suffers from both classical noise (generated due to random fluctuations of charged carriers within the electronic source devices) and quantum noise (arising due to entanglement, discord in transmission, decoherence, and randomness of the quantum world itself). The classical noise component is assumed to be Gaussian distributed with zero mean. The quantum noise component is assumed to be Poisson distributed. Starting from the assumption that both classical and quantum noise are additive, we formulate a hybrid noise model by deriving a joint Gaussian-Poisson distribution.
        


 
\vspace{-1mm}
\subsection{Signal Model}
{In our proposed communication system, the quantum link prepares a quantum state ensemble $\{p_X(x), \rho^x\}$ based on the classical input $X$ with probability $p_X(x)$, the corresponding density matrix $\rho^x$ and orthonormal basis $\{|x\rangle\}$ for a Hilbert space of dimension $|\chi|$, such that $\rho^x = \sum_x p_X(x)|x\rangle\langle x|$. We bring about correspondence between classical and quantum world as, $X \equiv \sum_{x \in \chi}x|x\rangle\langle x|$ and $\mathbb{E}_\rho[X] = \Tr\{X\rho\}$ where $\mathbb{E}[\cdot]$ represents the expectation of the classical variable $X$, and $\Tr\{\cdot\}$ represents the trace of the quantum counter-part.}

{The quantum states are transferred over a quantum channel $\mathcal{N}$ with unity-envelope to obtain output quantum states with orthonormal basis $\{|y\rangle\}$ which corresponds to the classical output $Y$ with distribution $p_Y(y)$. Classical information is retrieved from the quantum states by performing a measurement, the idea of which is captured in the positive operator-valued measure (POVM) \cite{17}, $\{\Lambda_y\}$. We can mathematically express the impact of the quantum channel $\mathcal{N}$ in terms of the Kraus operators, $\{\sqrt{p_{Y|X}(y|x)}|y\rangle\langle x|\}_{x,y}$, as,
\begin{align}
&\mathcal{N}(\rho^x) = \mathcal{N}\bigg(\sum_x p_X(x)|x\rangle\langle x|\bigg) \nonumber\\
&\quad = \sum_{x,y}\sqrt{p_{Y|X}(y|x)}|y\rangle\langle x|\bigg(\sum_{x'}p_X(x')|x'\rangle\langle x'|\bigg) \nonumber\\
&\quad\quad \times\sum_{x,y}\sqrt{p_{Y|X}(y|x)}|x\rangle\langle y|
\end{align}
Therefore, we can express,
\begin{align}
\mathcal{N}(\rho^x) &=\sum_{x,y}p_{Y|X}(y|x)p_X(x)|y\rangle\langle y|\nonumber\\
&\quad = \sum_y\bigg(\sum_x p_{Y|X}(y|x)p_X(x)\bigg)|y\rangle\langle y| \nonumber\\
&= \sum_y p_Y(y)|y\rangle\langle y|
\end{align}   
or we can write equivalently, $\sum_y p_Y(y)|y\rangle\langle y| = \mathcal{N}\big(\sum_x p_X(x)|x\rangle\langle x|\big)$. With correspondence to the classical world, we can write,} 
\begin{align}
{Y \equiv \mathcal{N}(X) + Z}
\end{align}
{where $Z$ is the noise added by the quantum channel and the classical system from which $X$ is generated. A unity-envelope channel refers to $\mathcal{N} = 1$ and therefore,} \begin{align}
{Y \equiv X + Z.}
\end{align}


{We define $Z=N_1 + N_2$, where $N_1$ is the discretely-distributed noise from the quantum world and $N_2$ is the continuously-distributed noise from the classical world. Here we are considering additive Poisson noise $N_{1}$ resulting from the environment over the quantum link. On the other hand, classical additive white Gaussian noise (AWGN) is added by the source that generates the classical information which is represented by $N_2$. The statistical distribution of the hybrid classical-quantum noise $Z$ is calculated by considering it as a convolution product of $N_{1}$ and $N_{2}$.}

The quantum noise $N_1$ follows a Poisson distribution \cite{1e,2e}  with  parameter $\lambda$, that is $N_1$ $\sim \mathcal{P}(\lambda) $, $\lambda \geq 0 , n \in  \mathbb{N}$; and the classical noise can be modelled by  Normal distribution with parameters $\mu$ and $\sigma$, where $\mu$  and $\sigma^2$ are the mean and variance of the distribution respectively,  $N_2 \sim \mathcal{N}$ ($\mu$,\,$\sigma^{2}$). 

As $N_1 \sim \mathcal{P}(\lambda)$, the probability mass function (pmf) of $N_1$  is given by,
 \begin{equation}
   f_{N_1}(n)=\frac{e^{-\lambda}\lambda^n}{n!} \label{eq}
 \end{equation}
\noindent where $\lambda \geq 0$ and $n$ $\in $ $\mathbb{N}$. Since $N_2 \sim \mathcal{N}$ ($\mu$,\,$\sigma^{2}$),  the probability density function (pdf) of $N_2$ can be expressed as,
 \begin{equation}
  f_{N_2}(t)=\frac{1}{\sigma\sqrt{2\pi}}e^{-\frac{1}{2}(\frac{t-\mu}{\sigma})^2 } \label{eq The pdf of Gaussian noise}
 \end{equation}

\noindent where $\mu$ and $\sigma$ are mean and standard deviation of the distribution.

Note that, $N_1 $ has a discrete distribution while $N_2 $ has a continuous distribution, and we aim to calculate their joint distribution. This can be done by expressing the pmf of the discrete Poisson distribution in terms of a pdf of a continuous distribution, and then calculating the joint distribution of two continuous distributions.
\vspace{-0.3 cm}
\subsection{Noise Model}
Consider a discrete random variable X with range $R_X = \{x_1, x_2,...\}$ and pmf $P_X(x_k)$. Note that the cumulative distribution function (c.d.f.) of $X$ can be written as 
   $F_X(x)= \sum_{\substack{
   \forall x_k \in R_X
   }} 
   P_X(x_k) u(x-x_k) $
 where $u(x)$ is the unit step function.
 
 We can write the pdf of $X$ by differentiating the c.d.f as follows:
  \begin{equation}
  \begin{split}
  f_X(x)& =\frac{dF_X(x)}{dx}  
            = \sum_{\substack{
              \forall x_k \in R_X
              }} 
              P_X(x_k)\frac{d}{dx}u(x-x_k)  \\
        & =\sum_{\substack{
             \forall x_k \in R_X
             }} 
             P_X(x_k)\delta(x-x_k) 
  \end{split}
 \end{equation} 
\noindent 
\\where $\delta(x)=\frac{d}{dx}u(x)$ is the Dirac delta function. The above expression is called the generalized pdf for the discrete random variable. 

In our case, $R_{N_1} =\mathbb{N} =\{1,2,3,\ldots,\infty\}$, $N_1$ is a discrete random variable (r.v.), $N_1 \sim \mathcal{P}(\lambda)$, and  the p.m.f is given by,
 \begin{equation}
     P_{N_1}(n)=\frac{e^{-\lambda}\lambda^n}{n!} 
     \label{eq pmf of N1}
 \end{equation}
 where $\lambda \geq 0 ,  n \in  \mathbb{N}$.
 
 The cumulative distribution function (c.d.f.) of the discrete r.v. $N_1 $ can be written as 
 \begin{equation}
   F_{N_1}(t)= \sum_{\substack{
    \forall n \in R_{N_1}
   }} 
   P_{N_1}(n) u(t-n) 
  \label{eq}
 \end{equation}
Now the pdf of the above function can be written as 
  \begin{equation}
  \begin{split}
      f_{N_1}(t) & =\frac{dF_{N_1}(t)}{dt} 
         = \sum_{\substack{
         \forall n \in R_{N_1}
          }}  P_{N_1}(t)\frac{d}{dt}u(t-n) \\
&=\sum_{\substack{
         \forall n \in R_{N_1}
         }} P_{N_1}(t)\delta(t-n) 
        =\sum_{\substack{
         \forall n \in R_{N_1}
         }}\frac{e^{-\lambda}\lambda^n}{n!}\delta(t-n)
        \label{eq The pdf of Poisson noise} 
     \end{split}
    \end{equation}
We know that if U is a discrete r.v. with pmf $p_U: \chi \rightarrow [0,1]$, and $\chi$ is a discrete set (may be countably infinite) then the r.v. U can be thought as a continuous r.v. with pdf 
    $f_U(u)= \sum_{\substack{
             \forall u_k \in \chi
             }} 
             p_U(u_k) \delta(u-u_k).$
  Now, if $V$ is a continuous r.v., and $W=U+V$ is a hybrid r.v. then the pdf of $W$ can be computed from pdfs of $U$ and $V$. Assuming $U$ and $V$ are independent r.v.s, the pdf of $W$ can be expressed as convolution product of pdfs  $f_U$ and $f_V$. Therefore,
    $f_W(w)= \sum_{\substack{
             \forall u_k \in \chi
             }} 
             p_U(u_k) f_V(w-u_k).$
For our case, from the expressions \eqref{eq The pdf of Gaussian noise} and \eqref{eq The pdf of Poisson noise} the pdf of  $Z$ can be written as, 
\begin{equation}
  \begin{split}
   f_Z(z) & 
       = \sum_{\substack{
            \forall n \in \mathbb{N}
             }} 
             \frac{e^{-\lambda}\lambda^n}{n!}\frac{1}{\sigma\sqrt{2\pi}}e^{-\frac{1}{2}(\frac{z-n-\mu}{\sigma})^2}
    \label{eq} 
    \end{split}
  \end{equation} 
We can then evaluate the differential entropy of r.v. $Z$  defined by 
    $ h(Z)=-\int_{\chi_{Z}}f_Z(z)log_{2} f_{Z}(z)\,dz $
where $\chi_{Z}$ is the support of $f_{Z}$, i.e., the set on which $f_{Z}$ is nonzero.\\

\vspace{-0.7 cm}
\subsection{Received Signal Model}      
    
If $P$ and $Q$ are two r.v.s with density functions $f_{P}(p)$ and $f_{Q}(q)$ defined $\forall$ $p$, $q$ in the respective domains, then the sum $R=P+Q$  is a r.v. with density  $f_{R}(r)$, where  $f_{R}$ is the convolution of  $f_{P}$  and $f_{Q}$. 

The convolution of two functions $f$ and $g$ can be defined as  $(f*g)(t):= \int_{-\infty}^{\infty} f(t-\tau)g(\tau) \,d\tau 
    =\int_{-\infty}^{\infty} g(t-\tau)f(\tau) \,d\tau $
    
If two r.v.s are related by $R=P+Q$, then pdf of $R$ can be written as the convolution product of the pdfs of $P$ and $Q$ as follows:
      $f_{R}(r):=\int_{-\infty}^{\infty} f_{PQ}(p,r-p) \,dp  $
Further, if P and Q are independent, then we can write:
     $ f_{PQ}(p,q)=f_{P}(p)f_{Q}(q) $
and the convolution product formula becomes 
    $f_{R}(r):=\int_{-\infty}^{\infty}f_{P}(p)f_{Q}(r-p) \,dp $
  
Since the distribution of the transmitted signal and the joint quantum-classical noise are independent of each other, their respective r.v.s are also independent. Therefore, we can write the pdf $f_{Y}$  of  $Y$ as the convolution of $f_{X}$  and $f_{Z}$ as follows:
\begin{equation}
     f_{Y}(y):=\int_{-\infty}^{\infty}f_{X}(x)f_{Z}(y-x) \,dx  \label{eq1}
\end{equation} 

{In absence of the knowledge on the exact distribution of the observable $X$, we approximate the transmit signal in terms of its point estimate (or mean), $\mu_X$, in order to choose a unique point in the signal parameter space which can reliably used to represent $X$. This point estimate $\mu_X$ is selected from the sample space generated over the range $R_{\mu_X} = [0, 2\pi]$.}

So from \eqref{eq1} using the convolution of $f_{X}$ and $f_{Z}$ we can approximate the joint pdf of $Y$ as :
 \begin{equation}
     f_{Y}(y) \approx \mu_{X} \int_{-\infty}^{\infty}f_{Z}(y-x) \,dx  
     \label{eq approximated the joint pdf of Y}
 \end{equation}
{where $f_Z(\cdot)$ is a function of pure-state qubits located on the surface of a unit Bloch sphere \cite{20}. Therefore, we can start by expressing $|x\rangle$ in terms of the elevation angle $\theta$ and azimuth angle $\phi$ of the sphere as,
$|x\rangle \equiv \binom{\cos \theta/2}{e^{i\phi \sin \theta/2}}$,
where $0 \leq \theta, \phi \leq 2\pi$. In order to reduce computation complexity, we restrict the representation of qubits as a function of a single parameter $\theta$, by defining $\phi$ as a function of $\theta$, such that $\phi = \psi(\theta)$.}

{This assumption is based on the the semi-classical concept of two-level quantum systems and Rabi oscillations. The Bloch sphere representation of a qubit can be modified in terms of the Rabi frequencies and two-level quantum states in the form of \cite{21}, $\theta = 2 \arccos[(\Omega/\Omega_g) \sin(\Omega_gt/2)]$ and $\phi = - \arcsin\bigg[\frac{\cos(\Omega_gt/2)}{\sqrt{1 - (\Omega/\Omega_g)^2} \sin^2(\Omega_gt/2)}\bigg]$ where, $\Omega$ is the Rabi frequency or the radian frequency at which electrons move from a higher to a lower level of energy, $\Omega_g$ is the generalized Rabi frequency or difference between the energy eigenvalues of the two energy states between which the transition takes place and $t$ is the time over which the transition between the two energy levels takes place. Therefore we can write,
\begin{align}
  \cos^2(\Omega_gt/2) = 1 - \frac{\Omega_g^2}{\Omega^2} \cos^2 (\theta/2)
\end{align}
and,
\begin{align}
   \phi &= -\arcsin\bigg[\frac{\cos(\Omega_gt/2)}{\sqrt{1 - (\Omega/\Omega_g)^2} \sin^2(\Omega_gt/2)}\bigg] \nonumber\\
   &= - \arcsin \bigg[\frac{\sqrt{1 - \frac{\Omega_g^2}{\Omega^2} \cos^2 (\theta/2)}}{\sqrt{1 - (\Omega/\Omega_g)^2} \frac{\Omega_g^2}{\Omega^2} \cos^2 (\theta/2)}\bigg] = \psi(\theta).
\end{align}}

In many cases, hybrid classical-quantum noise acts in a fixed direction. For example, it has been shown that for superconducting flux qubits, the chief noise source is flux noise \cite{22} which would put the noise only along the $z$-axis. This results in a physical phenomenon called `looping' on the Bloch sphere - that is the qubit after addition of noise returns to the neighborhood of its starting point after going around the sphere \cite{23}. It is due to the motion along lines of latitude on the sphere and only the $z$-component of the noise field produces this.

A mixture of quantum and classical noise, when represented on the Bloch sphere, rotates the angle $\theta = \arctan(\Delta/\varepsilon)$ so that the qubit energy eigenstates are aligned along the $z$-axis, where $\varepsilon$ and $\Delta$ are the energy difference and tunneling splitting respectively between two energy states. In this scenario, $\theta$ is referred to as the working point of the qubit.

So approximating $x$ by $\theta$ and varying $\theta$ over $0$ to $2\pi$, we have,
\begin{equation}
     f_{Y}(y)\approx\mu_{X} \int_{0}^{2\pi}f_{Z}(y-t) \,dt  
     \label{eq}
\end{equation}
 Finally, the differential entropy of the received signal $Y$ is defined by 
   $ h(Y)=-\int_{\chi_{Y}}f_Y(y)log f_{Y}(y)\,dy $
where $\chi_{Y}$ is the support of $f_{Y}$ .

\section{Capacity Calculation}


A quantum channel can carry both quantum and classical information. In this letter, we consider a unit-envelope quantum channel carrying classical information encoded into quantum states. Both our end users are classical in nature, while the actual communication takes place in the quantum domain. Our aim is to calculate how much classical information can be transmitted reliably per unit time over a quantum channel; hence the terminology `classical capacity of a quantum channel'.
 
\subsection{Capacity calculation of the Quantum Channel}

In classical communication theory, capacity of a power-constrained channel is given by $C=\max_{\substack{f(x) \\s.t. \mathbb{E}X^2] \leq P}}I(X;Y)$
where $f(x)$ is the distribution of $X$, $P$ represents the maximum transmit power and $\mathbb{E}[\cdot]$ denotes expectation of a random variable. Expanding $I(X;Y)$ in terms of differential entropies we have
   $I(X;Y) =h(Y)-h(Y|X)
           =h(Y)-h(X+Z|X)
          =h(Y)-h(Z|X) $.
where, $X$ and $Z$ are independent as the distribution of transmitted signal does not depend on the distribution of noise.

The quantity that governs how much classical information can be retrieved about random variable $X$ given $Y$, is the mutual information $I(X;Y)$, where $Y$ is the output of the quantum communication link. What kind of measurement has to be performed on $Y$ depends on what maximizes the information about $X$. The resulting quantity can be termed as the accessible information $\mathcal{I}_a(\varepsilon) \equiv \max_{\{\Lambda_y\}}I(X;Y)$, where $p_X(x)$ is the distribution of the input ensemble, $\varepsilon$ denotes the ensemble and $\Lambda_y$ is the POVM.

Since the `classical capacity of a quantum channel' is the maximum amount of classical information that can be transferred reliably over a quantum communication system in unit time, we represent it as the maximum mutual information between $X$ and $Y$ where the maximization is with respect to the input distribution; $C \equiv \max_{\{p_X(x), \rho^x\}} \mathcal{I}_a(\varepsilon) \equiv \max_{\{p_X(x), \rho^x\}} [\max_{\{\Lambda_y\}}I(X;Y)]$. In absence of the knowledge of the exact distribution of $X$, we resort to mean (point estimate) of $X$, thereby modifying the capacity definition to $C \equiv \max_{0 \leq \mu_X \leq 2\pi} \mathcal{I}_a(\varepsilon)$ where $\mu_X$ is the mean of the distribution of $X$.
   


\subsection{SNR Consideration}
{The SNR for our analysis can be expressed as, $\text{SNR}~= \frac{\mu_Y}{\sigma_Z }$
where $\mu_Y$ is the mean of the received signal and $\sigma_Z$ is the standard deviation of the noise. Consequently, $\mu_Y=\mathbb{E}[Y]=\int_{-\infty}^{\infty}yf_{Y}(y) \mathrm{d}y$ and
$\sigma_Z=\sqrt{\text{Var}(Z)}$ with $\text{Var}(Z) = \mathbb{E}[Z^2]-\mathbb{E}[Z]$, $\mathbb{E}[Z]=\int_{-\infty}^{\infty}zf_{Z}(z) \mathrm{d}z$ and $\mathbb{E}[Z^2]=\int_{-\infty}^{\infty}z^2f_{Z}(z) \mathrm{d}z$.} 
\section{Numerical Results}

We consider $\lambda = 5$ for the Poisson-distributed quantum noise, $\mu=0$ and $\sigma=15$ for the Gaussian-distributed classical noise, and $\mu_{X}=\pi$ for the expectation of transmitted signal.

\subsection{Characterization of joint quantum-classical noise:}
\label{AA}
\begin{figure}[t]
\centerline{\includegraphics[width = 1.05\columnwidth] 
{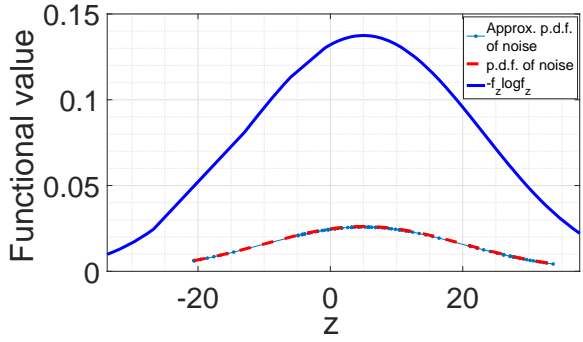}}
\caption{Comparative characterization of the pdf and the approximate pdf of the joint classical-quantum noise with the plot of $-f_Z(z)\log_{2} f_{Z}(z)$.}
\label{fig Characterization of joint noise}

\end{figure}

Consider the two following functions, 
\begin{equation}    
  f_Z(z)=  \sum_{n=1}^{\infty} 
           \frac{e^{-\lambda}\lambda^n}{n!}\frac{1}{\sigma\sqrt{2\pi}}e^{-\frac{1}{2}(\frac{z-n-\mu}{\sigma})^2 }
  \label{eq:pdfofZ}
\end{equation}
and 
 \begin{equation} 
  \tilde{f}_Z(z)=  \sum_{n=1}^{100} 
          \frac{e^{-\lambda}\lambda^n}{n!}\frac{1}{\sigma\sqrt{2\pi}}e^{-\frac{1}{2}(\frac{z-n-\mu}{\sigma})^2}
          \label{eq:pdfofZapprox}
\end{equation}
Theoretically, the density function in \eqref{eq:pdfofZ} is the actual pdf of $Z$ and the function in \eqref{eq:pdfofZapprox} is an approximated pdf of $Z$ shown in Fig.~\ref{fig Characterization of joint noise}.  Each simulation in Matlab randomly selects two sample spaces for Poisson (with parameter $\lambda = 5$) and Gaussian distribution (with parameters $\mu=0$ and $\sigma=15$), returning a joint sample space for $Z$. 
Let A and B be two sets, and $f:A \rightarrow B$ and $g:A \rightarrow B$ be two functions. Mathematically, two functions are identical if $f(a)=g(a)$ $\forall \ a\in A $. In our case, the domains of the functions  $f_{Z}$ and $\tilde{f}_Z$ are the same as they are the same sample space of $Z$, and $f_{Z}(z)= \tilde{f}_Z(z)$ $ \forall z\in Z $. Therefore the actual pdf of $Z$ in \eqref{eq:pdfofZ} can be well approximated by the function \eqref{eq:pdfofZapprox}, as can be seen in Fig.~\ref{fig Characterization of joint noise}. From here onward we will call $\tilde{f}_Z$ by ${f}_Z$.

In Fig.~\ref{fig Characterization of joint noise} we characterize the joint quantum-classical noise by plotting the function $-f_Z(z)\log_{2} f_{Z}(z)$ and $f_{Z}(z)$ where, the entropic measure of the joint quantum-classical noise is higher than the area under the noise pdf curve. 




\subsection{Mutual Information and Quantum Channel Capacity }
\begin{figure}[t]
\centerline{\includegraphics[width = 0.95\columnwidth]
{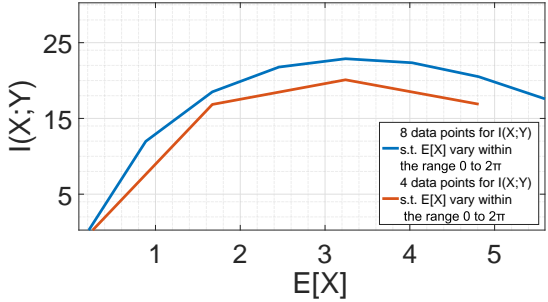}}
\caption{Mutual Information I(X;Y) vs. E[X] }
\label{fig Mutual Information I(X;Y) vs. E[X]}
\vspace{-0.3 cm}
\end{figure}

In Fig.~\ref{fig Mutual Information I(X;Y) vs. E[X]}, we vary the mutual information $I(X;Y)$ between the transmitted and received signal {over the expectation of} the transmit signal, $0 \leq \mu_{X} \leq 2\pi$. 
Approximating qubits as a single variable in 
we restricted the neighbourhood of the received qubits where the qubits can move only along one direction (longitudinal or latitudinal){on} the Bloch sphere.
\vspace{-0.5 cm}

\subsection{Capacity versus signal to noise ratio:}

\begin{figure}[t]
\centering
\subfloat[]{
   \includegraphics[width=0.47\columnwidth]{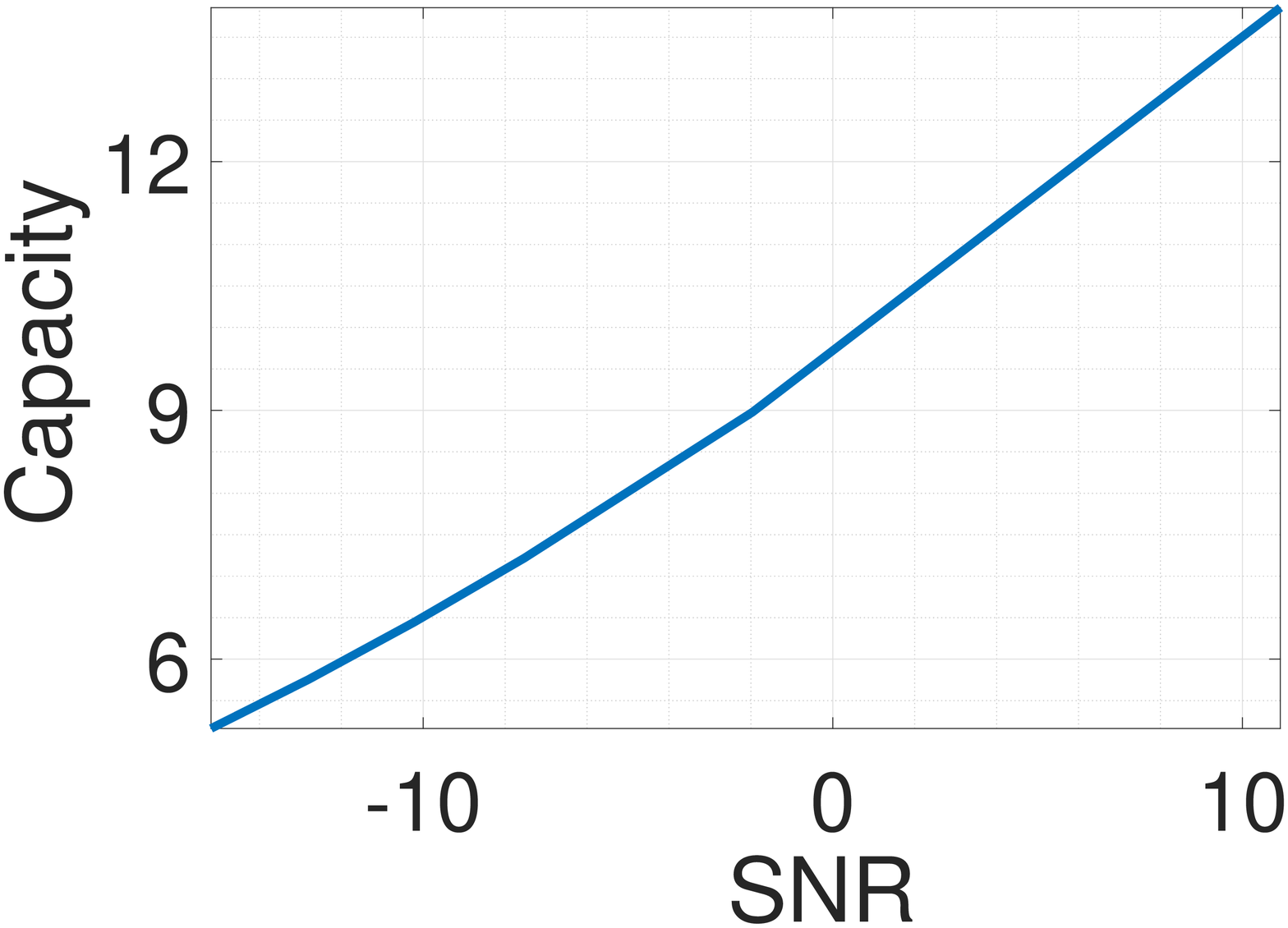}
  \label{fig_cap1}
}
\hfill
\subfloat[]{
   \includegraphics[width=0.47\columnwidth]{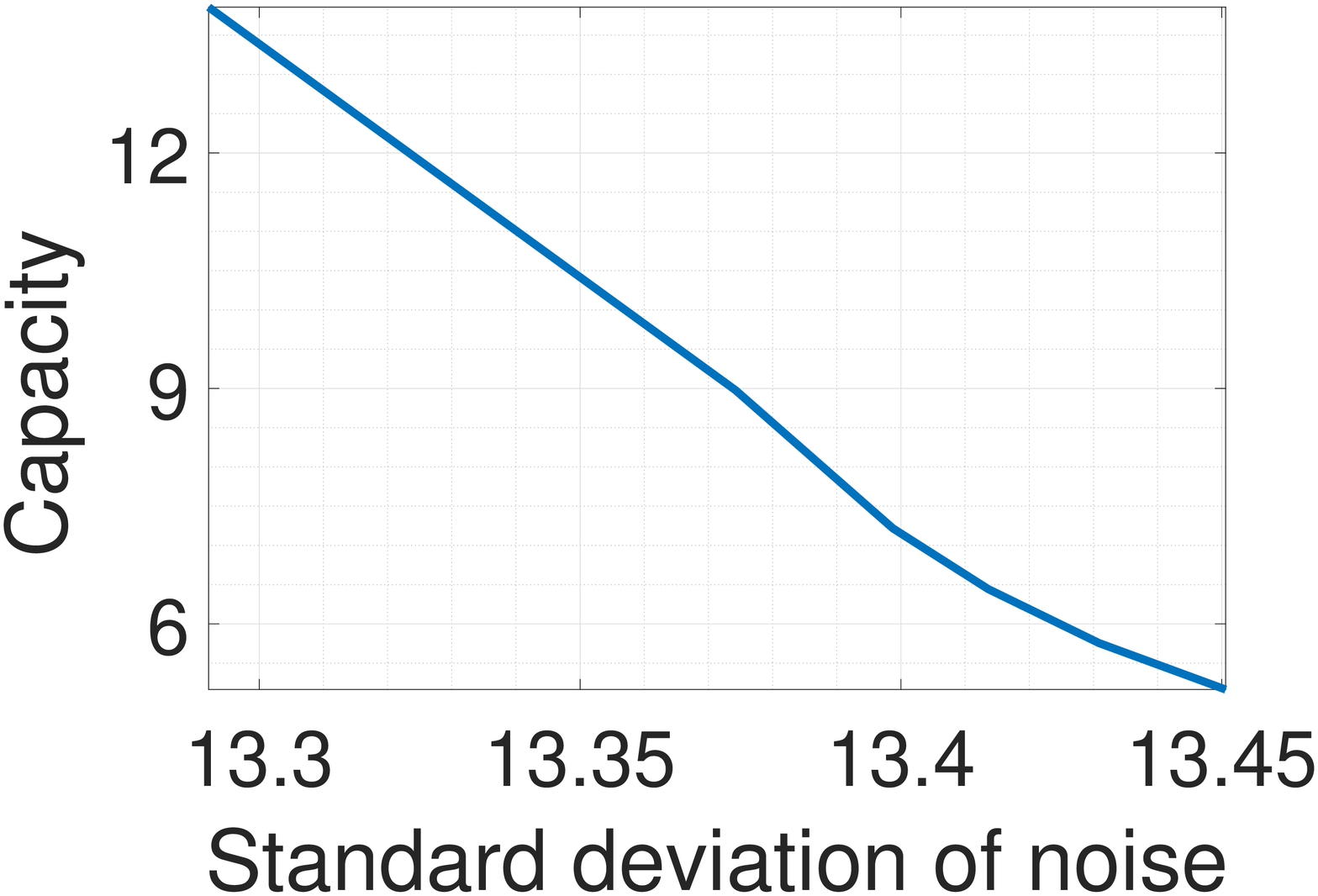}
 \label{fig_cap2}
}
\hfill
\caption{(a) Capacity with respect to SNR, and (b) capacity versus noise standard deviation}
\label{fig }
\end{figure}

In Fig.~\ref{fig_cap1} we plot the `classical capacity of a quantum channel' as a function of SNRs. We see that, capacity increases with SNR, as expected. Correspondingly, capacity decreases with increasing joint classical-quantum noise power thereby agreeing with the classical communication theory as demonstrated in Fig.~\ref{fig_cap2}.

\section{Concluding Remarks and Future Work}
In this letter we investigate the capacity that can be achieved over a unity-envelope quantum link carrying classical information, where the link is crippled with additive Gaussian-distributed classical and Poisson distributed quantum mechanical noise. Achievable capacity is calculated by maximizing the mutual information between classical source and destination and plotted as a function of SNRs, showing results in agreement with classical information theory. In future, we plan to extend our analysis to multiple channels with arbitrarily distributed transmit signals at the input of each channel. In this letter, we have restricted our analysis by considering the movements of qubits only in one direction. We are currently working towards relaxing this assumption by considering multiplicative noise, qubits as vector by designing their pdfs as vector-valued function; an interesting generalization which will be a contribution on its own.

\ifCLASSOPTIONcaptionsoff
  \newpage
\fi

\end{document}